\documentclass[pdflatex,sn-nature]{sn-jnl}

\usepackage{graphicx}%
\usepackage{multirow}%
\usepackage{amsmath,amssymb,amsfonts}%
\usepackage{amsthm}%
\usepackage{mathrsfs}%
\usepackage[title]{appendix}%
\usepackage{xcolor}%
\usepackage{textcomp}%
\usepackage{manyfoot}%
\usepackage{booktabs}%
\usepackage{algorithm}%
\usepackage{algorithmicx}%
\usepackage{algpseudocode}%
\usepackage{listings}%
\usepackage[caption=false,font=normalsize,labelfont=sf,textfont=sf]{subfig}
\usepackage{diagbox}
\usepackage{lmodern}

\begin{document}

\title[Article Title]{Meta-learning assisted robust control of universal quantum gates with uncertainties}


\author[1]{\fnm{Shihui} \sur{Zhang}}\equalcont{These authors contributed equally to this work.}

\author*[1]{\fnm{Zibo} \sur{Miao}}\equalcont{These authors contributed equally to this work.}\email{miaozibo@hit.edu.cn}

\author[2]{\fnm{Yu} \sur{Pan}}

\author[3]{\fnm{Sibo} \sur{Tao}}

\author*[4]{\fnm{Yu} \sur{Chen}}\equalcont{These authors contributed equally to this work.}\email{anschen1994@gmail.com}

\affil[1]{School of Mechanical Engineering and Automation, Harbin Institute of Technology, Shenzhen 518055, China}
\affil[3]{School of Mathematical Sciences, Fudan University, Handan Road 220, Shanghai 200433, China}
\affil[2]{College of Control Science and Engineering, Zhejiang University, Hangzhou, 310027, China}
\affil[4]{Mechanical and Automation Engineering, 
The Chinese University of Hong Kong, Shatin, Hong Kong SAR, China}

\abstract{Reliable quantum computing imposes higher requirements on the fidelity of quantum gates. However, because of the decoherence in quantum systems and certain delay or drift of control pulses, achieving the theoretical fidelity of controlled quantum gates in practice is always challenging. Inspired by meta-learning, in this paper we propose the meta-reinforcement learning quantum control algorithm (metaQctrl), which exhibits high robustness even if there exist uncertainties or disturbances in the system dynamics. The algorithm endows a quantum agent with awareness of dynamic environment through a two-layer learning framework. The concept of such a two-layer design facilitates separation of the environment and the decision-making process. This indeed allows the inner reinforcement learning network to focus on decision-making for specific optimization problems, while the outer meta-learning network adapts to varying environments and provides feedback to the inner network. Simulation examples regarding realization of the Hadamard gate, the $\pi/8$ gate, the phase gate, and the controlled-not (CNOT) gate demonstrate that metaQctrl manages to realize universal quantum gates robustly with uncertainties involved, outperforming its counterparts such as Gradient Ascent Pulse Engineering (GRAPE) and Proximal Policy Optimization (PPO), by employing fewer control pulses and obtaining higher fidelity values. Our results can thus contribute to the exploration of the quantum speed limit and the implementation of quantum circuits in the presence of system imperfections.}

\keywords{meta-learning, robust quantum control, quantum gates}



\maketitle
\section{Introduction}\label{sec1}

Recent theoretical and experimental studies have shown that quantum phenomena, such as entanglement and superposition, can be used to facilitate information science and technologies, including quantum computing, communications, and information storage, outperforming their classical counterparts \cite{Nielsen2011,XiaFeng2020,Khabat2016}. Quantum control is crucial to the successful realization of these leading technologies and also plays an important role in areas such as atomic physics and physical chemistry \cite{Chu2002,DongD2010}. 

Quantum control is usually implemented by manipulating the Hamiltonians of quantum systems. By applying a series of unitary transformations driven by the control Hamiltonians to the target system, the target system can be transferred to any desired state. Therefore, obtaining such a control sequence, especially the optimal control sequence, is a core problem that should be addressed by quantum control. A variety of control algorithms have been proposed in the past few decades, and have achieved excellent results in specific models. For example, the Gradient Ascent Pulse Engineering (GRAPE) algorithm \cite{Navin2005}, taking advantage of gradient-based optimization of a model-based target function, has seen successful applications in quantum optimal control.

However, in reality there are inevitably noises or inaccuracies in the information flow of quantum control systems, including but not limited to actuator output limitations \cite{DingHaiJin2019}, sensor measurement errors \cite{Hincks2015}, interactions with the environment and approximations of the physical models \cite{WangHanrui2022}. These factors can significantly influence the performance of quantum control systems that face the requirements of high precision. For instance, regarding the application of GRAPE, limited knowledge about the system dynamics and control Hamiltonians means that GRAPE may result in control deviations and even uncontrollability. Although earlier search algorithms such as gradient ascent (GA) exhibit some robustness, they require a lot of time to perform the optimization and are susceptible to falling into local optimal solutions \cite{Tsubouchi2008}. Therefore, designing an efficient algorithm to tackle uncertain dynamics of quantum systems remains challenging and meaningful.

Since machine learning (ML)  can efficiently and accurately complete classification or optimization tasks \cite{Shrestha2019}, the combination of ML and quantum engineering has been gradually explored \cite{WangZhe2022,ChenSamuelYenChi2022}. Owing to the advantages of ML, the difficulty in precisely modeling the dynamics of quantum systems can be relieved, thereby improving the control performance in various scenarios. In addition, reinforcement learning (RL) techniques in ML \cite{ZhangXiaoMing2019} have been adopted, including values-based algorithms \cite{Bukov2018} and policy gradient-based algorithms \cite{Moro2021}, to discover optimal quantum control strategies. 
RL typically consists of two basic components, namely, the environment and the agent. The environment can be established based on specific task requirements, such as games \cite{ShaoKun2019} and optimization problems \cite{PanZixiao2023}, while the agent is responsible for the decision-making process within the task and is supposed to learn the optimal policy by interacting with the environment. In particular, a series of RL algorithms have been specifically designed for the control of quantum gates through approaches such as the reconstruction of the reward function \cite{HuShouliang2022} and strategy optimization \cite{MaHailan2023}. 

However, existing results indicate that deep RL has at least two drawbacks compared to human beings. Training an intelligent agent requires a large amount of data, whereas humans can perform reasonably well in a novel environment with significantly smaller amounts of data. This limitation is prominent in quantum optimal control problems, as the exponential growth in the size of quantum states leads to a rapid decline in sample efficiency. On the other hand, deep RL specializes in a narrow task domain, while humans can effortlessly apply knowledge acquired from other scenarios to the current task. Consequently, improvements in generalization of RL algorithms have been considered in different tasks to overcome concomitant shortcomings \cite{Finn2017,Taylor2019,Khalid2023}. 

Inspired by deep meta-RL \cite{nagabandi2019,finnOnline2019,hospedalesSurvey2022,vettoruzzoReview2024}, in this paper, we propose a new algorithm which provides a generic and robust quantum agent that can adapt to different parametric settings in quantum systems. This algorithm significantly reduces the training time required for a quantum agent to handle diverse task environments and facilitates the transition of the algorithm among different control domains. Since quantum gates serve as fundamental operational elements in quantum computing \cite{AYuKitaev1997}, achieving high-fidelity quantum gates is of the utmost importance to ensure accuracy and efficiency in the vast majority of quantum technologies. The meta-reinforcement learning-assisted robust control algorithm proposed by us, metaQctrl, has been applied to design single-qubit and multi-qubit gates, with the purpose of obtaining ultrahigh-fidelity (of the order 99.99\%) in the presence of uncertainties or disturbances. For single-qubit quantum gates, compared to GA and GRAPE, our algorithm requires shorter control steps and exhibits smoother control curves. With disturbances or uncertainties involved, the fidelity of the quantum gate controlled by our algorithm is an order of magnitude higher than that by traditional algorithms (GA, GRAPE), as demonstrated in the simulation results. Furthermore, in the multi-qubit quantum gate control system with disturbances or uncertainties, our algorithm consistently outperforms the RL algorithm Proximal Policy Optimization (PPO) in terms of fidelity and control step length. Thus, it can be concluded that metaQctrl is capable of providing robust control strategies to adapt to complex environments. In addition, since metaQctrl requires fewer control pulses, our proposed algorithm can contribute to the exploration of the quantum speed limit when there are system imperfections \cite{Deffner2017,Taddei2013,ZhangMao2023}. 

The remainder of the paper is structured as follows. The mathematical model of quantum control systems with uncertainties is introduced in Section \ref{sec:rctrlmodel}, and how the framework of metaQctrl can be formulated is explained in Section \ref{sec:metaQctrl}. The robustness and superiority of the proposed algorithm are demonstrated in Section \ref{sec:rsoqgum} in the context of fast quantum gate control, where the level of robustness quantified by infidelity can reach the order of $10^{-4}$ subject to a wide range of deviations. And finally, Section \ref{sec:con} provides the concluding remarks.

\section{Robust control of quantum gates}\label{sec:rctrlmodel}

The evolution of a quantum system can be described by the following Schrödinger equation
\begin{equation}
{\left |\dot{\psi}(t)  \right \rangle}  = -iH\left | \psi(t)  \right \rangle = -i(H_{s}+\sum_{k=1}^{m}u_{k}(t)H_{k})\left | \psi(t)  \right \rangle.
\end{equation}
Here, \(\psi(t)\) represents the current quantum state of the system, $H_{s}$ denotes the inherent Hamiltonian of the system, and $H_{k}$ is the Hamiltonian generated by the external radio frequency (RF) field, with each RF field being independent and orthogonal. $u_{k}(t)$ denotes the amplitude of the control pulse. Furthermore, we assume that the target quantum system is governed by an internal Hamiltonian generated by spin-spin coupling. A RF field orthogonal to the spin \(x\)- and \(y\)-directions can then be used to manipulate the quantum system with the control pulse. We then can have the specific form of Hamiltonian as follows
\begin{equation}
H(\vec{u}(t) )=(1+\mu (t))[\sum_{j=1}^{n-1} Z_{j}Z_{j+1}+\sum_{j=1}^{n}(u_{x}^{j}X_{j}+u_{y}^{j}Y_{j})],
\end{equation}
where the operator $X_{j} = \mathrm {I}^{1}\otimes \cdots \otimes \mathrm {I}^{j-1}\otimes \sigma _{x}^{j}\otimes \mathrm {I}^{j+1}\otimes\cdots \otimes \mathrm {I}^{n}$ denotes the Pauli operator $\sigma _{x}$ applied on the $j$-th qubit in a system consisting of $n$ qubits ($\mathrm {I}$ denotes the identity operator). Similarly, $Y_{j}$ and $Z_{j}$ are defined corresponding to the Pauli operators $\sigma _{y}$ and $\sigma _{z}$ of the $j$-th qubit, respectively, and the control pulses are represented by $\vec{u}(t)$. This type of system model is widely applied in a variety of quantum systems, including superconducting qubits \cite{Barends2016} and trapped ions \cite{Saki2022}. In order to establish a universal control method that accounts for realistic imperfections, including dissipation, decoherence, and other uncertainties, the parametric disturbance $\mu (t)$ is introduced into our control model.  Please note that there may be multiple disturbances or uncertainties appearing in the system (i.e. $\mu_1 (t)$, $\mu_2 (t)$, $\cdots$), and in this section we take $\mu (t)$ to illustrate the control scheme without loss of generality. To further improve the adaptive capacity of the algorithm, we assume that $\mu (t)$ obeys a truncated Gaussian distribution with the variance $\eta$ being uncertain and the mean being $0$. That is, we consider $\mu (t)\sim \mathrm{clip} [\mathbb{N}(0,\eta^{2}),-1,1]$. 

We can use $T$ to denote the end time and $N$ to denote the steps of the control pulses. And then the quantum state of the target system can be calculated by 
\begin{equation}
\left | \psi (t) \right \rangle =\mathcal{D}e^{-i\int_{0}^{T}d\tau H(\tau)}  \left | \psi (0) \right \rangle =U(T)\left | \psi (0) \right \rangle,
\end{equation}
where $\mathcal{D}$ denotes the Dyson time-ordering operator. In particular, the quantum gate $U(T)$ can be given by $U(T)=\prod_{j=1}^{N} U_{j} (\Delta t)$ with $U_{j}(\Delta t) =e^{ -iH(\vec{u}(j\Delta t) )\Delta t}$ and $\Delta t=T/N$. The control goal is to drive the controlled quantum gate towards the desired quantum gate $U_{f} $ that makes $\left | \psi (t) \right \rangle =U_{f} \left | \psi (0) \right \rangle $. Our aim is to discover the optimal control scheme that enables the controlled quantum gate to transit from the initial state to the target state with greater precision and in less time. 

The precision of a controlled quantum gate can be quantified by the fidelity \cite{YuHaixu2023,Zahedinejad2015}
\begin{equation}
\mathcal{F} (U(T),U_{f})=\left | \frac{\mathrm {Tr}\left [ U_{f}^{\dagger }U(T)\right ]}{2^{n}}  \right |^{2}, 
\end{equation}
with the range of $\mathcal{F}$ being $\left [ 0 ,1\right ]$. The closer the fidelity is to 1, the higher the precision of the quantum gate. The objective function of the quantum gate control problem can then be defined by
\begin{equation}
\mathcal{U} ^{*} =\mathrm {arg\ }\underset{\mathcal{U}}{\mathrm {max}}\int \mathrm{d}\mu (t)\mathcal{F}(U(T),U_{f}).   
\end{equation}
The definition of $\mathcal{U}$ should be emphasized as it significantly deviates from traditional quantum control problems. Previous studies, in general, focus on identifying the optimal control $\vec{u}^*_{\mu (t)}$ in undisturbed environments ($\mu (t)\equiv 0$) or in fixed disturbed environments ($\mu (t)$ subject to a specific distribution), to make $\mathcal{F}(U(T), U_f)$  as close to $1$ as possible. In contrast, our goal is to seek a generic mapping $\mathcal{U}^*: \mu (t) \to \vec{u}^*_{\mu (t)}$ rather than solving specific cases. This pursuit is more challenging than obtaining solutions for a given $\mu (t)$. For example, if an algorithm has been developed to determine $\vec{u}^*_{\mu (t)}$, such as GRAPE and GA, one would have to redesign or train the algorithm to adapt to the task in a new environment, which is indeed very time consuming. However, our algorithm, which will be explained in detail in the following section, intelligently reduces the transition time between different disturbed environments. And therefore, the robustness can be enhanced.

\section{The framework of metaQctrl}\label{sec:metaQctrl}

The decision-making process in the framework of metaQctrl is derived from RL, based on the Markov Decision Process (MDP). MDP can be defined by a 5-tuple $\left ( S,A,P,R,\gamma  \right )$, where $S$ represents the state space, $A$ denotes the action space, $P$ is the transfer function providing the probability of transferring from a given state to a new one after taking an action, $R$ is the reward function mapping the result of an action taken in a specific state to a value, and $\gamma$ represents the discount factor.

\begin{figure}[htbp]
\centering
\includegraphics[width=3.5in]{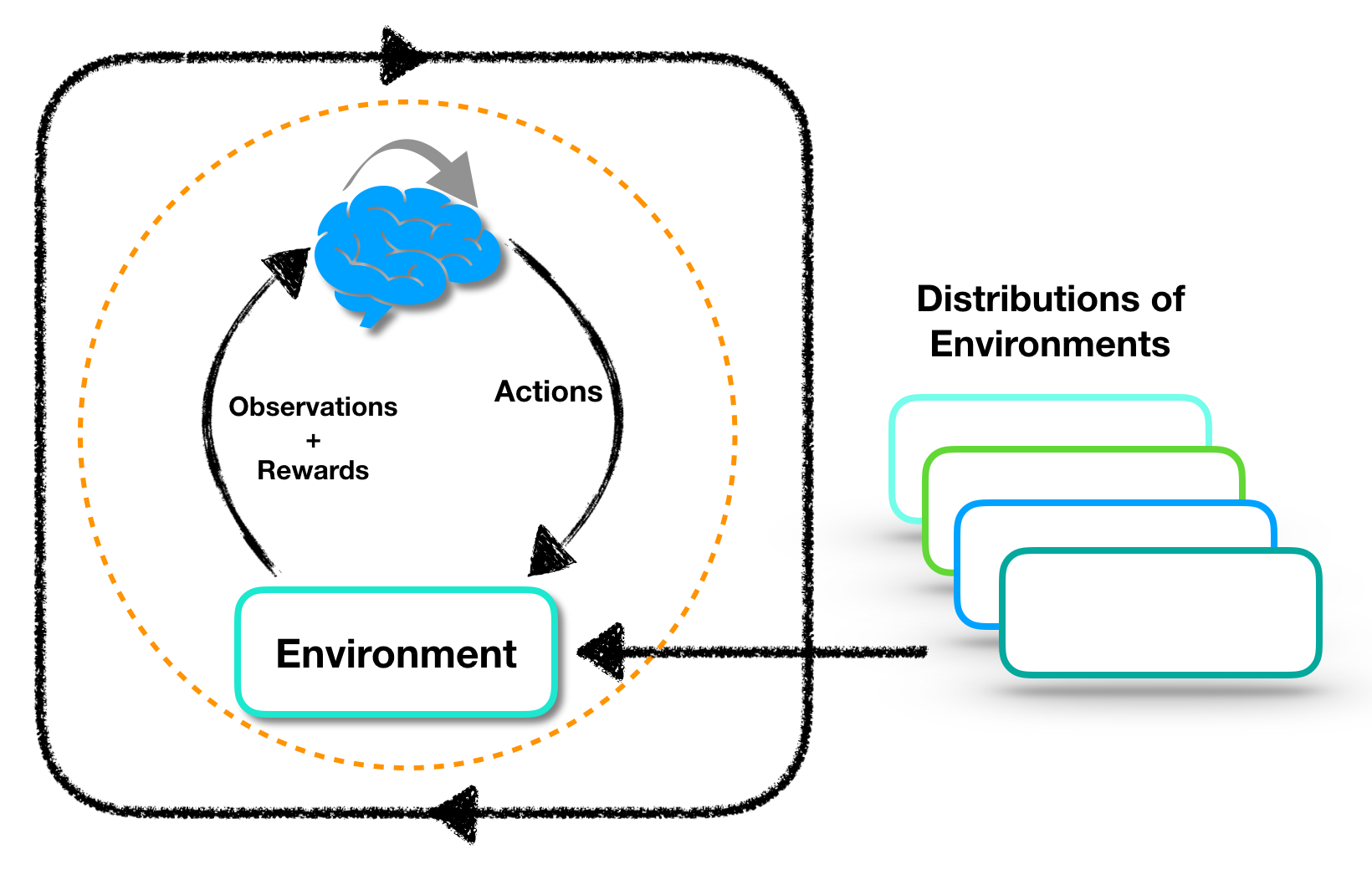}
\caption{ Illustration of how a quantum agent can be constructed. The yellow loop represents the general setup of RL. Within this loop, the agent engages with a specific RL environment to acquire the optimal strategy.  The environment of metaQctrl encompasses multiple RL environments adhering to a specific distribution. Classifying these environments into distinct tasks, a quantum agent is supposed to autonomously adapt their algorithms to accommodate changes in environments.}
\label{fig_1}
\end{figure}

As shown by the yellow loop in Figure \ref{fig_1}, RL can be viewed as an interactive process between an agent and the corresponding external environment.
Based on the current state of the environment $s_{t} $ and the associated policy function, the agent (denoted by the blue brain) takes the action $a_t$ at the time $t$, leading to three consequences. Firstly, the agent receives a reward $r_t$, and then the state of the environment changes from $s_t$ to $s_{t+1}$. Finally, the agent receives a new observation $s_{t+1}$. Through continuous interactions with the environment, the agent has access to a substantial amount of decision-making data. The data in turn enable further optimization of the policy function.

Throughout this process, the agent aims to find a policy $\pi(a|s)$ to optimize the total expected reward defined as follows
\begin{equation}
    J(\theta) = \sum_{s\in S} d^{\pi}(s)V^{\pi}(s) = \sum_{s\in S}d^{\pi}(s) \sum_{a\in A}\pi_{\theta}(a|s) Q^{\pi}(s,a),
\end{equation}
where,
\begin{align}
&V^{\pi}(s) = \mathbb{E}_{a\sim \pi}\left[\sum_{k\ge 0}\gamma^k r_{t+k}|s_t=s\right], \\ 
&Q^{\pi}(s,a) = \mathbb{E}\left[\sum_{k\ge 0}\gamma^k r_{t+k}|s_t=s, a_t=a\right],
\end{align}
and $d^{\pi}(s)$ is the stationary distribution of the Markov chain concerning $\pi_{\theta}$ (i.e., on policy state distribution under $\pi$), mathematically defined by $d^{\pi}(s) = \lim_{t\to\infty} P(s_t=s|s_0, \pi)$. The policy-based RL algorithm here is applied to optimize the policy $\pi$, approximated by a neural network $\pi_\theta$, to maximize the reward function $J(\theta)$ ($\theta$ represents the parameter of the neural network). A well-known rule, referred to as the policy gradient theorem, can be used to update the reward function as follows
\begin{equation}
    \aligned
    \nabla_\theta J(\theta)  & = \nabla_\theta \sum_{s\in S}d^{\pi}(s) \sum_{a\in A}\pi_{\theta}(a|s) Q^{\pi}(s,a) \\ & \propto \sum_{s\in S}d^{\pi}(s) \sum_{a\in A} Q^{\pi}(s,a) \nabla_{\theta}\pi_{\theta}(a|s).
    \endaligned
\end{equation}

However, in general, RL algorithms face challenges in deriving the optimal solution for the objective as discussed in Section \ref{sec:rctrlmodel} due to their specificity to the environments. Once the environment changes, the originally learned optimal strategy becomes obsolete and necessitates retraining for the new environment, which is highly impractical for quantum systems. Therefore, in this paper, we introduce meta-learning to established RL algorithms, since meta-learning can help acquire knowledge quickly and adapt to new task scenarios based on existing knowledge. This process can also be perceived as maximizing the sensitivity of the new task loss function to the model parameters. In fact, even small alterations in model parameters can lead to significant enhancements in the loss function when the sensitivity is high. The intelligent quantum agent can be constructed by training the model in various tasks, as illustrated in Fig. \ref{fig_1}. Especially for the control objective $\mathcal{U} ^{*} $, metaQctrl can contribute to increasing robustness in the presence of system disturbances or uncertainties.

The associated model parameters can be defined as follows. The state $s_{t}$ is given by $s_{t}=\{ \Re ([U]_{00} ),\dots , \Re ([U]_{2^{n}2^{n} }), $ $\Im ([U]_{00}),\dots ,\Im ([U]_{2^{n}2^{n} } ) \}$, where $[U]_{jk}$ represents the element at the intersection of the $j$-th row and $k$-th column of the unitary operator $U(t)$ that describes the controlled quantum gate at time $t$. $\Re (\cdot )$ and $\Im (\cdot )$ denote the real and imaginary parts of an operator, respectively. The action $a_{t} =\left \{ u_{x}^{1},\dots,u_{x}^{n}, u_{y}^{1},\dots,u_{y}^{n}  \right \} $, and each $u_k^j$ is bounded within the range $[u_{\mathrm{min}},u_{\mathrm{max}}]$. The reward $r_{t} $ is set as a piecewise function related to $\mathcal{F} (U(T),U_{f})$, namely
\begin{equation}
r_t= \begin{cases}
500(1-t/T), & \mathcal{F} >1-\varepsilon  \\ 
10(1-t/T), &0.98<\mathcal{F}\le1-\varepsilon \\
1-t/T, &0.9<\mathcal{F}\le0.98 \\
-(1-\mathcal{F}), &\mathcal{F}\le 0.9
\end{cases}
\label{eq:epscc}
\end{equation} 
where $\varepsilon$ defines the convergence criterion. By mapping the fidelity of quantum gates to rewards, our framework effectively transforms the control of quantum gates into a solvable task using RL algorithms.
When fidelity is not high enough, an increase in fidelity leads to an increase in the corresponding reward, thus motivating the agent to quickly approach the target state. When fidelity is already high, segmented rewards are used to encourage exploratory behavior aimed at discovering strategies with even higher fidelity. In addition to promoting exploration, we also incorporate a penalty coefficient for step length to expedite convergence and prevent excessive exploration.

\begin{figure}[htbp]
\centering
\includegraphics[width=5in]{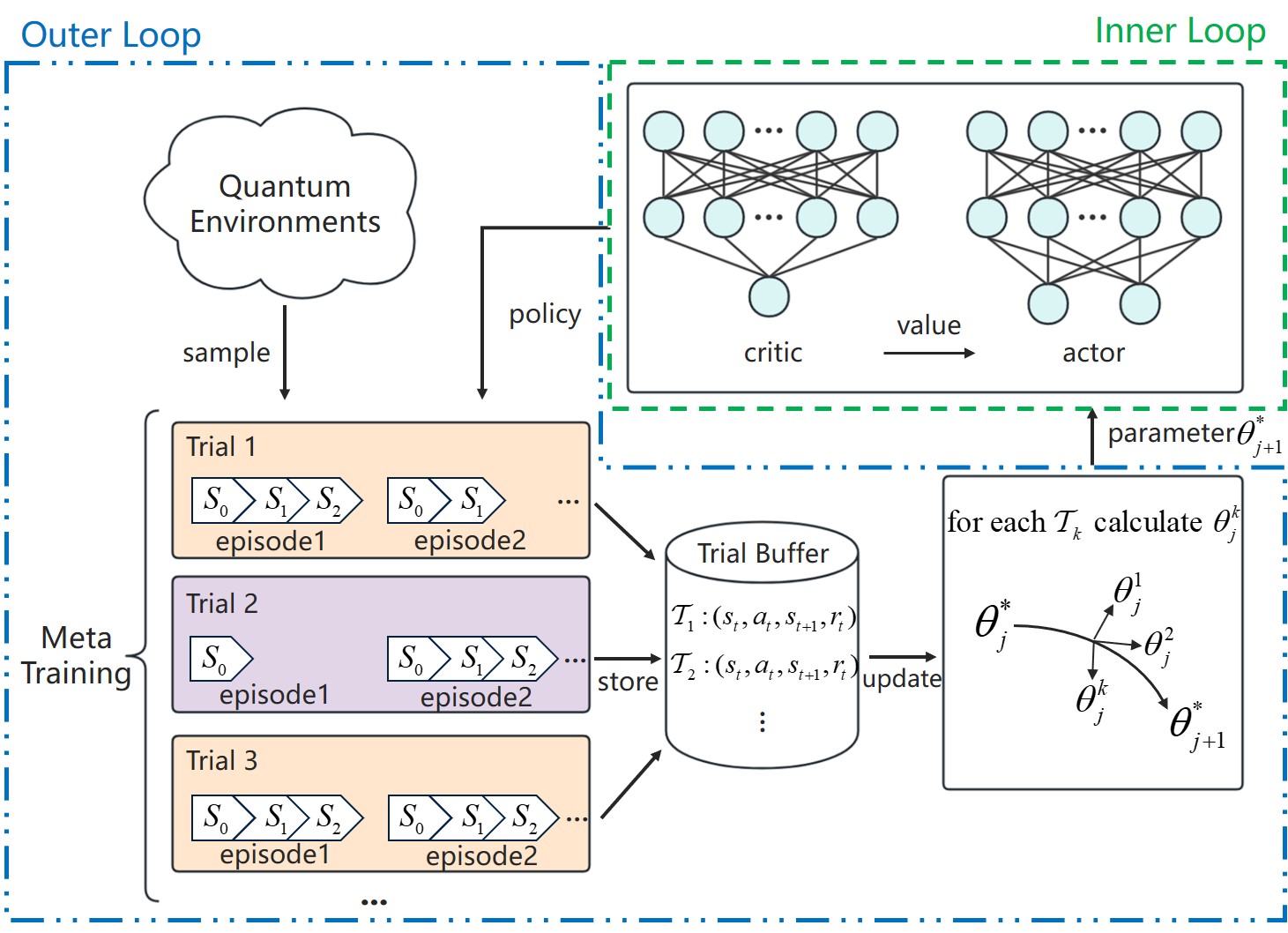}
\caption{The framework of metaQctrl. The training process of the inner loop resembles general RL for a specific task, while  meta-learning is embodied in the outer loop. This outer loop partitions the results obtained from sampling the quantum environment into different tasks and subsequently delivers them to the inner loop for processing. Subsequently, data acquired from each inner loop is collected, culminating in utilizing this collective information to update the parameters of the quantum agent.}
\label{fig_2}
\end{figure}

The metaQctrl algorithm, as illustrated in Fig. \ref{fig_2}, comprises an inner loop and an outer loop. The outer loop represents a meta-learning process that maintains a global policy parameter $\theta_{j} ^{*} $. It acquires a set of tasks $\mathcal{T}$ by sampling the quantum environment $\mathcal{P}$. Each task, denoted $\mathcal{T} _{k}$, can be viewed as an individual RL environment. By receiving each task $\mathcal{T} _{k}$ from the outer loop, the inner loop performs an independent RL with its current policy parameter $\theta_{j} ^{*} $ determined by the outer loop. It then executes the MDP and collects state-reward pairs to update its own policy parameters accordingly. To accomplish the procedures, the inner loop employs an actor-critic network-based policy improvement method to update network parameters as follows,
\begin{equation}
\theta _{j}^{k} =\theta _{j}^{*} +\nabla _{\theta _{j}^{*}}\log{\pi_{\theta _{j}^{*}}(s,a)\mathcal{A}_{\pi}(s,a)} .
\end{equation}
Here, $\mathcal{A}_{\pi}(s,a)$ denotes the advantage function whose value can be estimated by the value network. Once the inner loop is updated to obtain a new parameter \( \theta _{j}^{k} \), the MDP is immediately reexecuted with the updated policy parameter. 
The newly obtained state-reward pairs can then be used to perform the following update process \cite{Finn2017}
\begin{equation}
\theta _{j+1}^{*}=\theta _{j}^{*}-\beta \nabla _{\theta _{j}^{*}}\sum_{k}\mathcal{L} _{\mathcal{T}_{k}}(f_{\theta _{j}^{k}}).
\end{equation}
The parameter $\beta$ defines the update rate of meta-learning. $f_{\theta _{j}^{k}}$ denotes the neural network of inner loop with the updated policy parameter $\theta _{j}^{k}$, and $\nabla _{\theta _{j}^{*}}\mathcal{L} _{\mathcal{T}_{k}}(\cdot)$ denotes the gradient of the returns associated with the updated policy. 
The core idea of our framework is to take the magnitude of the loss value as weights, with the update direction of each subtask used as the basic direction for updating the global initialization parameters. It is worth noting that by taking advantage of the loss value of the new strategy as weights instead of directly summing gradients, it mitigates the risk of algorithmic convergence towards local optima. Through multiple iterations, the outer loop can acquire prior knowledge about dynamic environments in quantum systems from various tasks, thereby achieving adaptability. Furthermore, the outer loop samples disturbance distributions and performs meta-updates, which can be seen as a process of estimating the disturbance. Consequently, the inner loop can adaptively adjust control signals in varying noisy environments, thus enhancing robust performance against disturbance.

Inspired by the policy approximation methods and the Kullback-Leibler divergence formula \cite{Schulman2017}, we optimize the update method for meta-policy parameters to better measure the policy distance and effectively control the policy update step size. Namely
\begin{equation}
\theta _{j+1}^{*} \gets \mathrm {arg}\ \underset{\theta }{\mathrm {max}} {\textstyle \sum_{\mathcal{T}_{k}\sim \mathcal{P}}}\mathcal{L}_{\theta _{j}^{k}}^{\rm{Clip}}(\theta),
\end{equation}
where
\begin{align}
\mathcal{L} _{\theta _{j}^{*}}^{\rm{Clip}}(\theta ) = {\mathbb{E}_{\pi _{k}}}\left [ \sum_{t = 0}^{T}\left [ \mathrm {min}(c_{t}(\theta)\hat{\mathcal{A}}_{t}^{\pi_{k}},\mathrm {clip}(c_{t}(\theta),1-\epsilon,1+\epsilon )\hat{\mathcal{A}}_{t}^{\pi_{k}})    \right ]   \right ],
\label{eq:umfmpp}
\end{align} 
and $\hat{\mathcal{A}} _{t}^{\pi _{k}} =Q^{\pi _{k}} (s,a)-V^{\pi _{k}}(s)$. The difference between the new and the old policies is given by $c_{t}(\theta )=\frac{\pi_{\theta }(a_{t}|s_{t})}{\pi_{\theta_{\rm{old}}}(a_{t}|s_{t})}$. When the new and old policies are identical, $c_{t}(\theta )=1$. Both $\min(\cdot)$ and the clip$(\cdot)$ in Eq. \eqref{eq:umfmpp} are used to ensure that the step size of the meta-updating process takes a value within the predefined range $[1-\varepsilon ,1+\varepsilon ]$.

As illustrated in Algorithm \ref{alg1}, whenever a control task is started, it requires collecting a new batch of data online, which is then used to update the network parameters.
Using the prior knowledge obtained during pretraining, the algorithm can efficiently identify the optimal policy parameters tailored to the current noisy conditions. Upon completion of the training stage, the system can output a control sequence according to the control task. Despite relying on a sampling training process, this algorithm benefits from a more thoroughly trained model parameter, enabling the online training process to remain fast. Remarkably, its average runtime is significantly less than that of traditional algorithms such as GRAPE and GA, which will be explained in detail in the following section.

\begin{algorithm}[H]
\caption{metaQctrl}\label{alg:alg1}
\hspace*{0.02in}{\bf Input:}
Target quantum gate $U_{f}$, distribution of quantum environment $\mathcal{P}$, convergence criterion $\varepsilon $, length of training steps $\mathcal{J} $\\
\hspace*{0.02in}{\bf Output:} 
 Optimal policy $\theta ^{*} $
\begin{algorithmic}[1] 
\State {Initialize quantum gate $U(0)$, meta-hyperparameter $\theta _{0}^{*}$,  training steps $j$, trial buffer $B$}
\State {Build a actor-critic network with weights $\theta _{0}^{*}$}
\While{$j<\mathcal{J} $}
\State{sample batch of tasks $\mathcal{T} _{k}$ from $\mathcal{P}$}
\While{$\mathcal{T} _{k}$}
\State{initialize the inner actor-critic network with $\theta_{j} ^{*} $}
\State{run the inner actor-critic network to get $n$ episodes}
\State{compute advantage estimates for each steps}
\State{compute $\theta _{j}^{k} =\theta _{j}^{*} +\nabla _{\theta _{j}^{*}}\log{\pi_{\theta _{j}^{*}}(s,a)\mathcal{A}_{\pi}(s,a)}$}
\State{run the actor-critic network with $\theta _{j}^{k}$ again and store the episodes into $B$}
\EndWhile
\State{update $\theta _{j+1}^{*} \gets \mathrm {arg}\ \underset{\theta }{\mathrm {max}} {\textstyle \sum_{\mathcal{T}_{k}\sim \mathcal{P}}}\mathcal{L}_{\theta _{j}^{k}}^{\rm{Clip}}(\theta)$ with episodes in $B$}
\State{$j=j+1$}
\EndWhile
\end{algorithmic}
\label{alg1}
\end{algorithm}

\section{Robust control of quantum gates using metaQctrl}
\label{sec:rsoqgum}

To demonstrate the features and strengths of metaQctrl, in this section, we discuss robust control of single- and multi-qubit quantum gates in the presence of uncertainty. We numerically compare our proposed algorithm with GRAPE, GA, and PPO within the same system setup in the simulation, with the assistance of OpenAI's Gymnasium and QUTIP \cite{JRJohansson2013}.

\subsection{Robust control of a single-qubit quantum gate}\label{subsec:rcosqg}

Considering robust control of a single-qubit quantum gate, the system Hamiltonian can be described by
\begin{equation}
H(\vec{u}(t))=(1+\mu (t))(\sigma _{z}+u_{x}(t)\sigma_{x} +u_{y}(t)\sigma_{y}),
\end{equation}
where the control variables $u_{x}(t)$ and $u_{y}(t)$ are restricted in the interval $\left [ -5,5 \right ]$. The control horizon is $T=1.6$ and the maximum steps of the control pulses are $N=40$. The uncertainty involved in the system, $\mu(t)$, is assumed to obey the Gaussian distribution $\mathbb{N} (0,\eta^{2} )$, with the standard deviation $\eta \in \left [ 0,1 \right ]$. The control objective is to transfer the quantum gate from the initial state $U(0)$ to the final state $U(T)$, with the end time $T$. The fidelity $\mathcal{F}(U(T),U_{f}) >1-\varepsilon$ must be guaranteed as prescribed in Eq.~\eqref{eq:epscc}. Specifically, we take $U(0)=\mathrm {I}=\begin{pmatrix} 1 & 0\\ 0 & 1\end{pmatrix}$, $U_{f} =U_{\rm{H}}=\frac{1}{\sqrt{2}}\begin{pmatrix} 1 & 1\\ 1 & -1\end{pmatrix}$ with $U_{\rm{H}}$ the Hadamard gate widely used in quantum computing and the convergence criterion $\varepsilon =10^{-4}$.  

\begin{figure}[H]
\centering
\subfloat[GRAPE]{\includegraphics[width=2.5in]{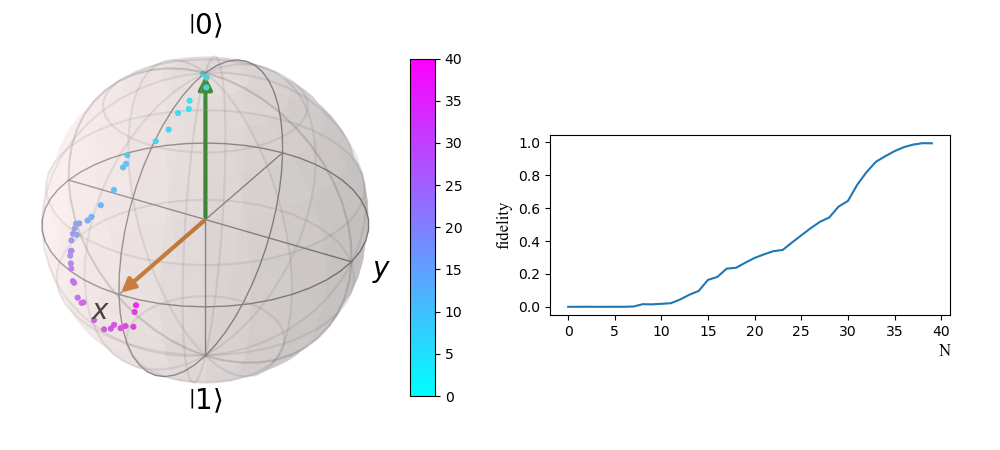}%
\label{fig_3_1}}
\hfil
\subfloat[GA]{\includegraphics[width=2.5in]{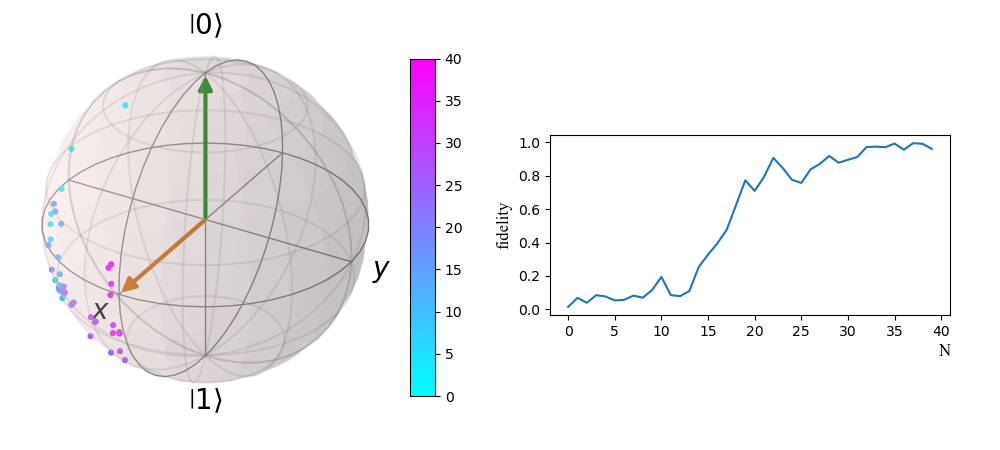}%
\label{fig_3_2}}
\hfil
\subfloat[PPO]{\includegraphics[width=2.5in]{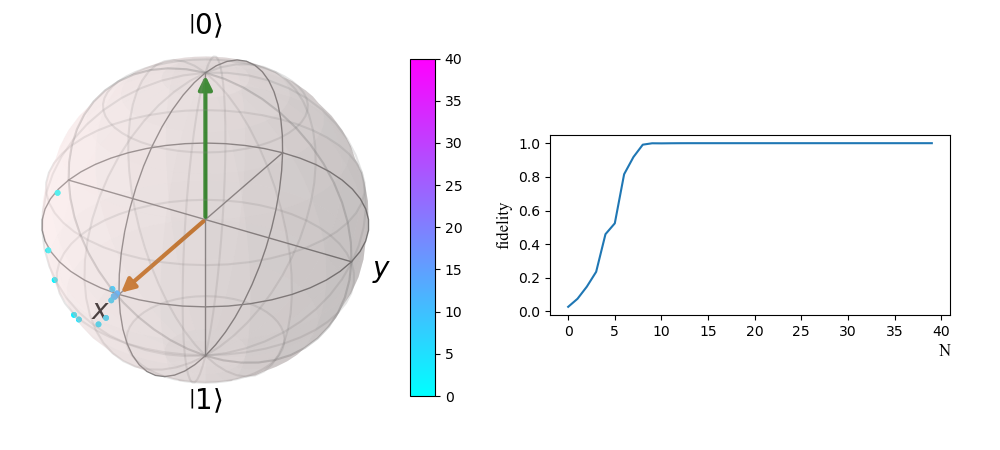}%
\label{fig_3_3}}
\hfil
\subfloat[metaQctrl]{\includegraphics[width=2.5in]{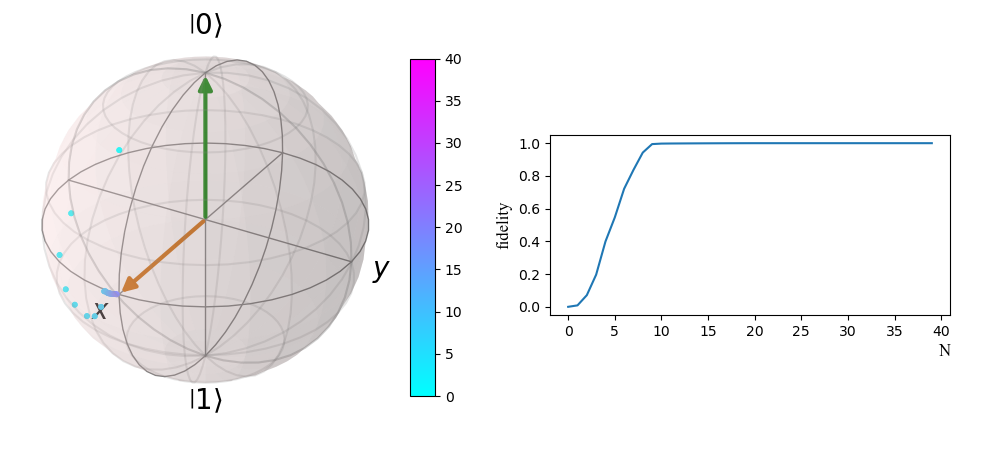}%
\label{fig_3_4}}
\caption{The system evolution of (a )GRAPE, (b) GA, (c) PPO, (d) metaQctrl in realization of single-qubit Hadamard gate with $\eta =0.3$. On the left side of each graph, the Bloch sphere is utilized to visually show the state trajectory over time, since changes in quantum gates can be observed through changes in quantum states. The initial and target quantum states are dots on the Bloch sphere, centered by the green and orange arrows, respectively. The blue dots denoting the outcome of the control action correspond to earlier steps, while the purple dots indicate steps closer to completion. If a dot reaches the orange arrow before reaching the maximum number of control steps, it implies the successful completion of the control task. On the right side of each graph, numerical results demonstrate how the fidelity of the quantum gate changes as the number of control steps increases.}
\label{fig_3}
\end{figure}

The control trajectories of four algorithms (GRAPE, GA, PPO, metaQctrl) corresponding to $\eta = 0.3$ are depicted in Fig. \ref{fig_3}. 
It can be clearly seen that only metaQctrl and PPO successfully complete the control task in the presence of disturbance or uncertainty, while GRAPE and GA fail to meet the fidelity accuracy requirements within the maximum control step limit. Although GRAPE and GA exhaust all allocated control steps, this does not imply that our control step length $N$ is unreasonable, since metaQctrl and PPO achieve their control objectives in fewer than $N/2$ steps. It can be concluded that metaQctrl and PPO, which combine efficient exploration with accuracy, outperform GRAPE and GA in realizing global optimal control. 
In fact, metaQctrl and PPO adopt a strategic exploration approach by boldly advancing when far from the target and making incremental adjustments as they approach, thus preventing them from missing the target and falling into local optima.

\begin{figure}[H]
\centering
\subfloat[]{\includegraphics[width=2.1in]{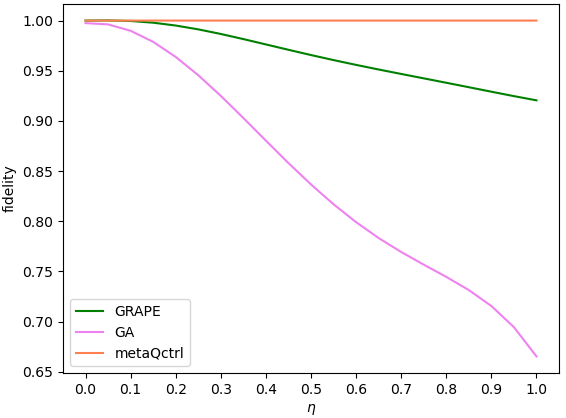}%
\label{fig_4_1}}
\hfil
\subfloat[]{\includegraphics[width=2.1in]{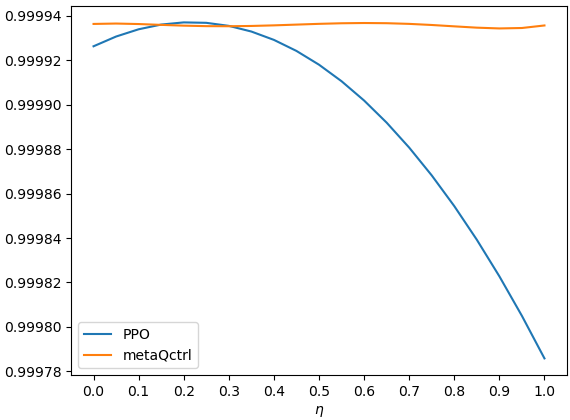}%
\label{fig_4_2}}
\hfil
\subfloat[]{\includegraphics[width=2.5in]{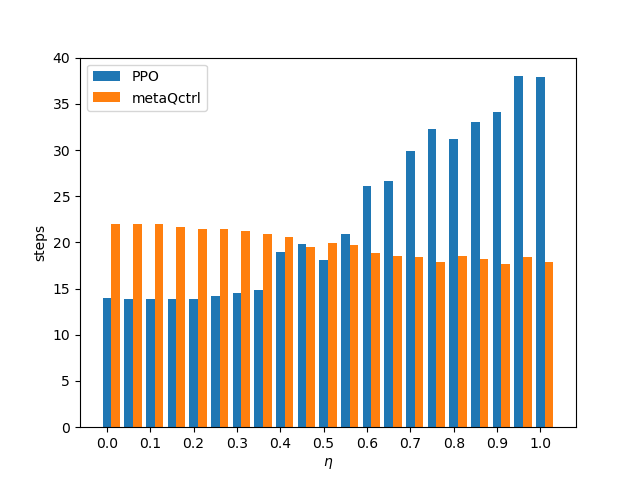}%
\label{fig_4_3}}
\subfloat[]{\includegraphics[width=2.35in]{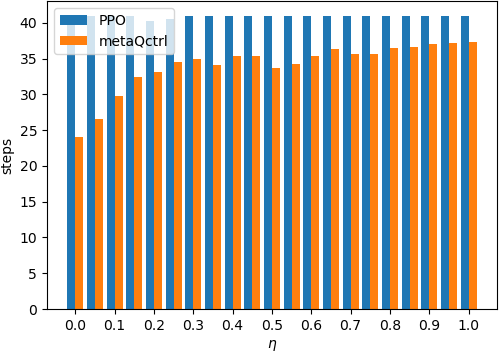}%
\label{fig_4_4}}
\caption{The control performance comparison among GRAPE, GA, PPO and metaQctrl in realization of single-qubit Hadamard gate with varying $\eta$, where (a) the averaged maximum fidelity of metaQctrl, GRAPE and GA when $\varepsilon = 10^{-4}$; (b) the averaged maximum fidelity of metaQctrl and PPO when $\varepsilon = 10^{-4}$; (c) the averaged number of control pulses required for both metaQctrl and PPO to reach the maximum when $\varepsilon = 10^{-4}$; (d) the averaged number of control pulses required for both metaQctrl and PPO to reach the maximum when $\varepsilon = 10^{-6}$, are plotted.}
\label{fig_4}
\end{figure}

In addition, the maximum fidelity achieved by four algorithms with varying \(\eta\) is evaluated in Fig. \ref{fig_4}. Here, \(\eta\) uniformly ranges from $0$ to $1$ with a step size of $0.05$. The simulation results represent the average values obtained from the $100$ Monte Carlo tests. PPO is trained with the disturbance parameter \(\eta = 0.3\), while metaQctrl can handle the disturbance distribution $\mathcal{P}$ with $\eta \sim \mathbb{U}(0,1)$. In particular, the convergence criterion $\varepsilon$ is set to $10^{-4}$ in Figs. \ref{fig_4_1},\ref{fig_4_2} and \ref{fig_4_3}, while the convergence criterion $\varepsilon$ is set to $10^{-6}$ in Fig. \ref{fig_4_4}.
The fidelity values obtained by metaQctrl compared to GRAPE and GA are plotted in Fig. \ref{fig_4_1}. It is evident that metaQctrl exhibits significantly enhanced robustness compared to traditional algorithms. The differences between traditional methods and metaQctrl are subtle in uncertainty-free scenarios. However, as the intensity of the disturbance increases, so does the performance gap. Even in the extreme case (\(\eta = 1.0\)), metaQctrl achieves a maximum fidelity $10\%$ higher than GRAPE and $30\%$ higher than GA. The control performances of metaQctrl and PPO in terms of maximum fidelity are compared in Fig. \ref{fig_4_2}. 
It can be seen that metaQctrl excels in maintaining robust control performance across a broader range of disturbances, leveraging its unique meta-reinforcement learning framework. Fig. \ref{fig_4_3} compares the control steps of metaQctrl and PPO when $\varepsilon = 10^{-4}$. It can be observed from Fig. \ref{fig_4_3} that the exploration intensity of PPO increases in order to handle the system uncertainties, resulting in a rapid escalation of its total control steps. In contrast, the control steps for metaQctrl remain relatively stable across varying disturbance distributions and even exhibit a slight decreasing trend as $\eta$ increases. However, as shown in Fig. \ref{fig_4_4}, the number of control steps for metaQctrl increases as $\eta$ increases when $\varepsilon = 10^{-6}$. This is due to the fact that the number of control steps required by metaQctrl is closely related to the choice of the convergence criterion $\varepsilon$ and indeed the scale of the target neighborhood relative to the disturbances. When $\varepsilon = 10^{-4}$, the scale of the target neighborhood relative to the disturbances is sufficiently large, and consequently, the trajectory of the system is unlikely to deviate from the target neighborhood. Since metaQctrl aims to avoid the oscillation of the control trajectory in the target neighborhood, it accelerates convergence during the later stages of control, and thus the number of control steps decreases slightly as $\eta$ increases in this case. In contrast, when $\varepsilon = 10^{-6}$, the scale of the target neighborhood relative to the disturbances is not large enough. More pronounced oscillations near the target region during later stages of control increase the number of control steps as $\eta$ increases.

In order to further explore the robustness of metaQctrl against PPO, we take into account multiple uncertainties or disturbances in the Hamiltonian, which gives
\begin{align}
H(\vec{u}(t))=(1+\mu_{0} (t))\sigma _{z}+ (1+\mu_{u} (t))(u_{x}(t)\sigma_{x} +u_{y}(t)\sigma_{y}).
\end{align}
Here, $\mu_{0}(t)$ and $\mu_{u}(t)$ obey the Gaussian distributions $\mathbb{N} (0,\eta_0^{2} )$ with standard deviation $\eta_{0} \sim \mathbb{U}(0,1)$ and $\mathbb{N}(0,\eta_u^{2} )$ with standard deviation $\eta_{u} \sim \mathbb{U}(0,1)$, respectively. The control objective is to transfer the quantum gate from $U(0) = \mathrm {I}$ to $U(T)= U_{\rm{H}}=\frac{1}{\sqrt{2}}\begin{pmatrix} 1 & 1\\ 1 & -1\end{pmatrix}$ (the Hadamard gate) or $U(T)= U_{\rm{\pi/8}}=\begin{pmatrix} 1 & 0\\ 0 & e^{i\pi/4}\end{pmatrix}$ (the $\pi/8$ gate) or $U(T)= U_{\rm{S}}=\begin{pmatrix} 1 & 0\\ 0 & i\end{pmatrix}$ (the phase gate), with the end time $T=1.6$ and the maximum steps of the control pulses $N=40$.

\begin{figure}[!htbp]
\centering
\subfloat[]{\includegraphics[width=3.8in]{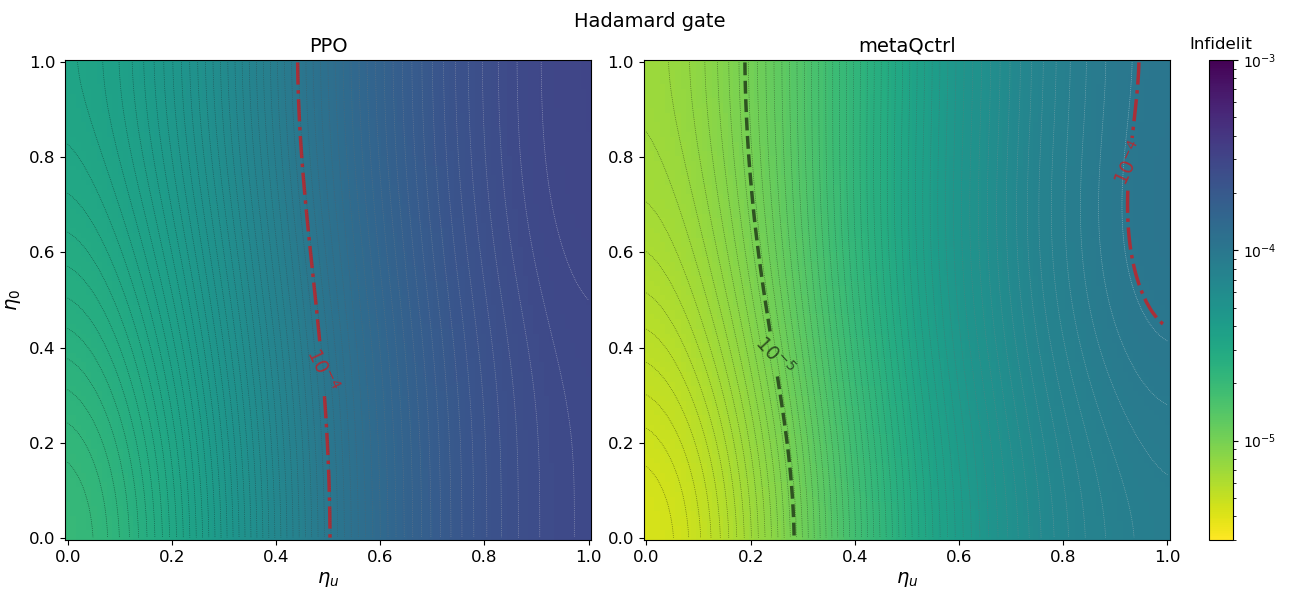}%
\label{fig_5_1}}
\hfil
\subfloat[]{\includegraphics[width=3.8in]{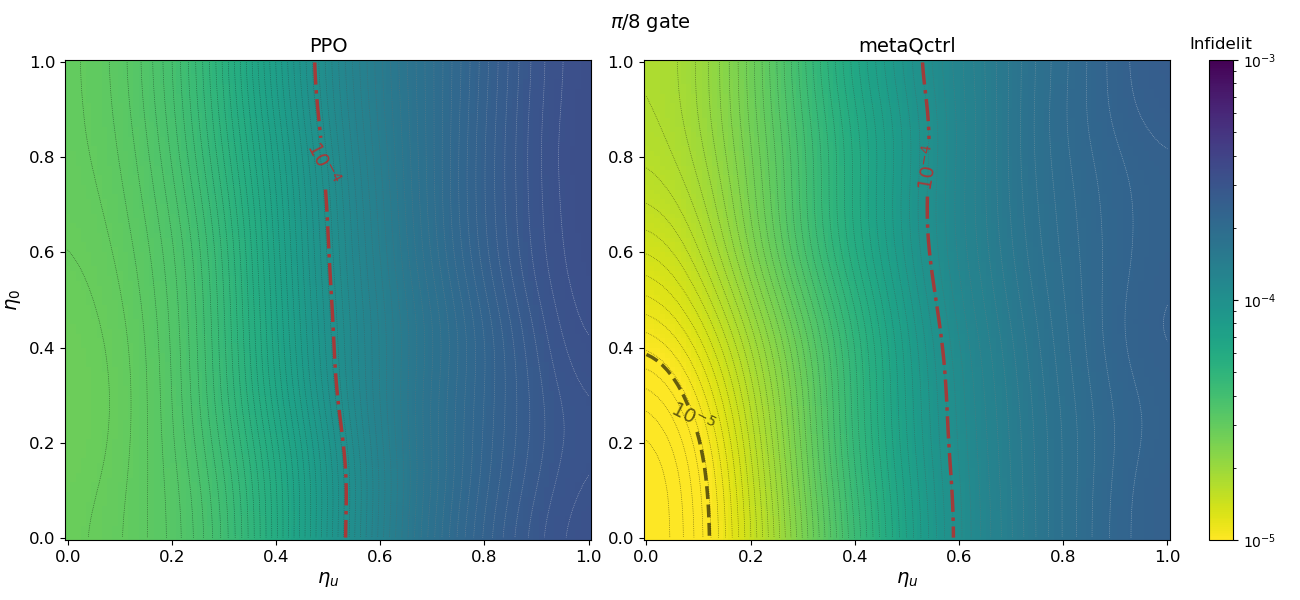}%
\label{fig_5_2}}
\hfil
\subfloat[]{\includegraphics[width=3.8in]{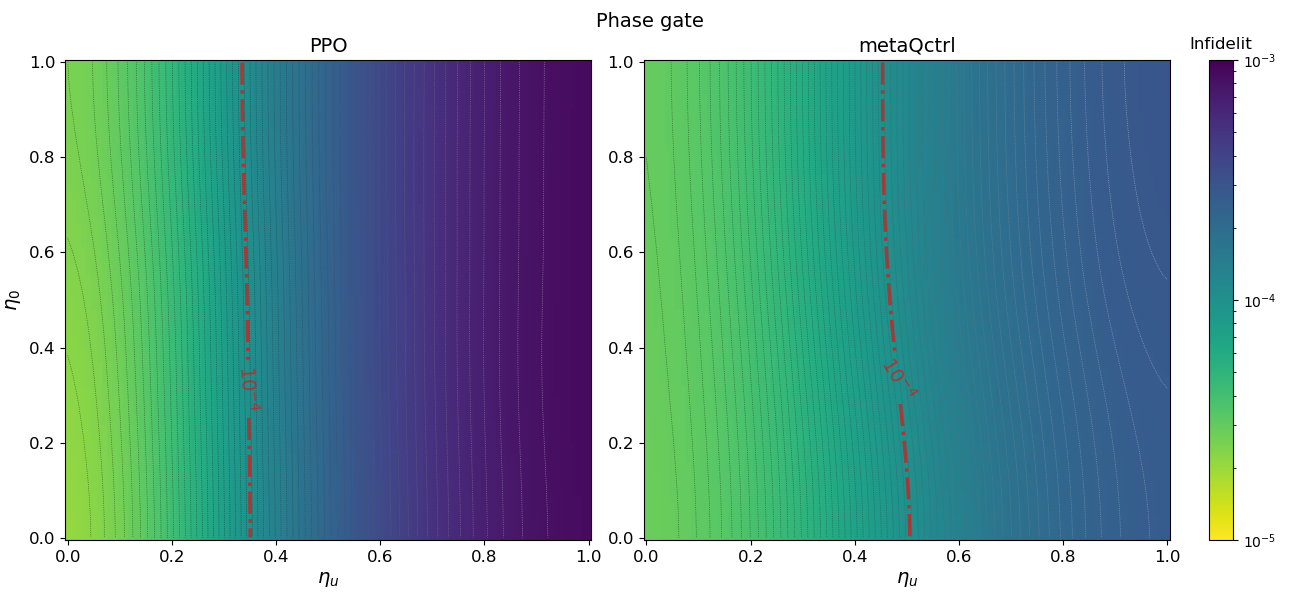}%
\label{fig_5_3}}
\caption{The control performance of metaQctrl, compared to that of PPO, in realization of single-qubit (a) Hadamard gate, (b) $\pi/8$ gate, and (c) phase gate, with the uncertainties $\mu_{0}(t)$ and $\mu_{u}(t)$ involved. The black dashed and red dash-dotted boundary lines correspond to the values of infidelity at $10^{-5}$ and $10^{-4}$ respectively.}
\label{fig_5}
\end{figure}

\begin{table}[ht]
    \renewcommand{\arraystretch}{1.5}
    \centering
    \begin{tabular}{|c|c|c|c|}
        \hline
        \diagbox{Algorithms}{Quantum gates} & Hadamard gate & $\pi/8$ gate & phase gate \\ \hline
        PPO & \diagbox{$0\%$}{\textcolor{red}{$47.4\%$}} & \diagbox{$0\%$}{\textcolor{red}{$50.7\%$}} & \diagbox{$0\%$}{\textcolor{red}{$34.3\%$}} \\ \hline
        metaQctrl & \diagbox{$23.2\%$}{\textcolor{red}{$96.5\%$}} & \diagbox{$3.1\%$}{\textcolor{red}{$55.9\%$}} & \diagbox{$0\%$}{\textcolor{red}{$47.3\%$}} \\ \hline
    \end{tabular}
    \caption{The ratios of the areas corresponding to the key infidelity thresholds ($10^{-4}$ and $10^{-5}$), relative to the entire space enclosed by $\eta_0$ and $\eta_u$ for metaQctrl compared to PPO, in realization of Hadamard gate, $\pi/8$ gate and phase gate. The values above the diagonal lines give the area covered ratios when the infidelity is not greater than $10^{-4}$, whereas the values below the diagonal lines give the area covered ratios when the infidelity is not greater than $10^{-5}$.}
    \label{tab_InfidelityArea1}
\end{table}

As shown in Fig.~\ref{fig_5}, metaQctrl obviously outperforms PPO by presenting stronger robustness quantified by infidelity in the presence of multiple uncertainties. It can be seen that the infidelity values can reach $10^{-5}$ or even smaller ones (e.g. for the Hadamard gate) using metaQctrl when the standard deviation $\eta_0$ ($\eta_u$) is not large. Moreover, metaQctrl allows for a wider range of deviations than PPO does, where fidelity values greater than 99.99\% can be obtained even if the system is subject to multiple uncertainties. Table \ref{tab_InfidelityArea1}, along with Fig.~\ref{fig_5}, provides ratios of the areas corresponding to the key infidelity thresholds ($10^{-4}$ and $10^{-5}$), relative to the entire space enclosed by $\eta_0$ and $\eta_u$, for metaQctrl and PPO, respectively.

\subsection{Robust control of a two-qubit quantum gate}\label{subsec:rcotqg}

We now consider robust control of a two-qubit quantum gate, with the Hamiltonian of the system given by
\begin{equation}
H(\vec{u}(t))=(1+\mu (t))(Z_{1}Z_{2}  +\sum_{j=1}^{2}(u_{x}^{j}(t)X_{j}+u_{y}^{j}(t)Y_{j}) ),
\end{equation}
where the control variables $u_{x}^{j}(t),u_{y}^{j}(t)\in\left [-5,5\right ]$ ($j\in \left \{ 1,2 \right \}$), and control horizon is $T=2.0$. The maximum step of the control pulses is $N=50$, with the quantum gate initiated at $U(0)=\mathrm {I}$. In this scenario, the desired quantum gate for control is $U_{f} =U_{\rm{CNOT}}$ (there is no local z phase error correction involved), with $U_{\rm{CNOT}}$ the controlled-NOT (CNOT) gate, and the convergence criterion is characterized by $\varepsilon =10^{-3}$. The CNOT gate is of the following form
\begin{equation}
U_{\rm{CNOT}} =\begin{pmatrix}  1&  0&  0& 0\\  0&  1&  0& 0\\  0&  0&  0& 1\\  0&  0&  1& 0\end{pmatrix}.
\end{equation}

\begin{figure}[!htbp]
\centering
\subfloat[]{\includegraphics[width=2.1in]{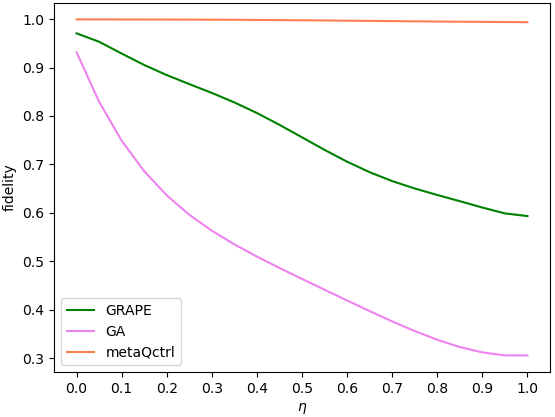}%
\label{fig_6_1}}
\hfil
\subfloat[]{\includegraphics[width=2.1in]{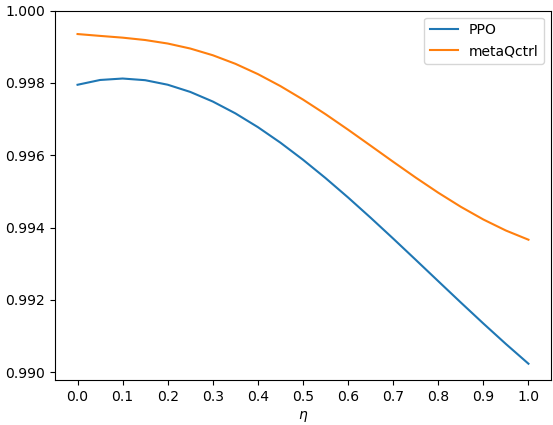}%
\label{fig_6_2}}
\hfil
\subfloat[]{\includegraphics[width=2.5in]{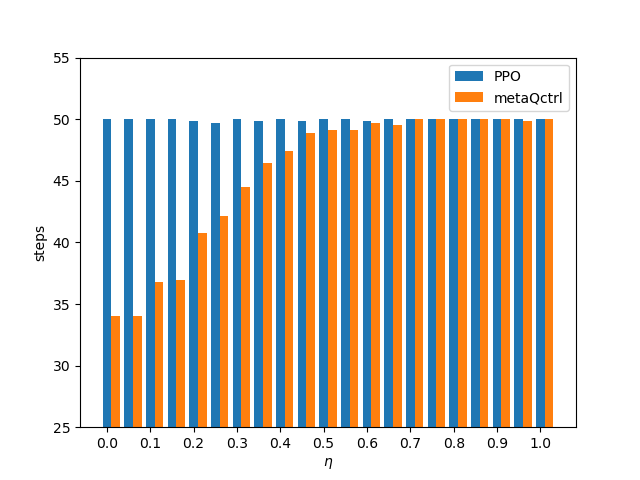}%
\label{fig_6_3}}
\caption{The control performance comparison among GRAPE, GA, PPO and metaQctrl in realization of CNOT gate with varying $\eta$, where (a) the averaged maximum fidelity of metaQctrl, GRAPE and GA; (b) the averaged maximum fidelity of metaQctrl and PPO; (c) the averaged number of control pulses required for both metaQctrl and PPO to reach the maximum, are plotted.}
\label{fig_6}
\end{figure}

The maximum fidelity achieved by four algorithms (GRAPE, GA, PPO, metaQctrl) with varying \(\eta\) is evaluated in Fig. \ref{fig_6}. As shown in Fig. \ref{fig_6_1}, the control performance of metaQctrl is compared with that of GRAPE and GA. Traditional algorithms (GRAPE and GA) are increasingly sensitive to uncertainty as \(\eta\) increases in a two-qubit system. Consequently, the fidelity of the quantum gates they control drops quickly below $0.9$, failing to meet the requirements of actual control tasks. However, metaQctrl can provide infidelity of the order $10^{-4}$. 
Fig. \ref{fig_6_2} shows the maximum fidelity values between metaQctrl and PPO. It can be seen that metaQctrl and PPO are also affected by the increase in the number of qubits, resulting in slightly lower fidelity for two-qubit systems compared to single-qubit systems. In addition, metaQctrl shows a more gradual downward trend and achieves higher fidelity than PPO with respect to different values of $\eta$. This indicates that the control algorithm proposed in this paper is indeed more robust. It can be understood that metaQctrl has learned to grasp dynamic environmental information from a higher dimension and can quickly adjust its strategy according to changes in the environment, through training with the meta-reinforcement learning framework. 
By comparing the total control steps in Fig. \ref{fig_6_3}, we observe that while PPO exhausts all available control steps, metaQctrl has the ability to search for effective control strategies, with shorter control sequences when \(\eta\) is small. 



\section{CONCLUSION AND DISCUSSION}\label{sec:con}

Drawing inspiration from meta-learning, the metaQctrl algorithm proposed in this paper is aimed at enhancing the control performance of quantum gates in the presence of system imperfections. To be specific, in the framework of metaQctrl, we incorporate a meta-learning outer loop apart from the RL inner loop and employ an alternating training approach. This two-layer structure significantly enhances the algorithm's robustness against environmental changes. The meta-learning outer loop enables extraction of specific tasks in the presence of uncertainties. By acquiring knowledge from various tasks, the inner loop requires only minimal new data to determine optimal control strategies for novel tasks, thereby realizing robust control of quantum gates with disturbances or uncertainties involved. Furthermore, in the context of simulation examples, we have evaluated the performance of our proposed algorithm with respect to robust control of single-qubit quantum gates and the CNOT gate that constitute the universal gate set. The results and analyses show that metaQctrl obviously outperforms GRAPE or GA, and that metaQctrl can achieve improved robustness with fewer control pulses compared to PPO. It is thus promising that metaQctrl can help to discover the quantum speed limit. Additionally, it is worth highlighting that metaQctrl can be straightforwardly applied to quantum metrology or communication related to quantum state manipulation with system imperfections, not limited to the area of quantum computing such as robust realization of quantum circuits. Future work may include generalizing our method to the more complicated quantum systems and taking into account totally time-varying disturbances.



\bibliography{reference}

\end{document}